# Intel® SGX Enabled Key Manager Service with OpenStack Barbican

**Somnath Chakrabarti, Brandon Baker, Mona Vij**

Intel Labs
2111 NE 25th Ave, Hillsboro, OR 97124
{somnath.chakrabarti, brandon.baker, mona.vij}@intel.com

**ABSTRACT**
Protecting data in the cloud continues to gain in importance, with encryption being used to achieve the desired data protection. While there is desire to use encryption, various cloud components do not want to deal with key management, which points to a strong need for a separate key management system. OpenStack Barbican is a platform developed by the OpenStack community aimed at providing cryptographic functions useful for all environments, including large ephemeral clouds. Barbican exposes REST APIs designed for the secure storage, provisioning and management of secrets such as passwords, encryption keys, and X.509 certificates, and supports plugins for a variety of crypto solutions in the backend. Crypto plugins store secrets as encrypted blobs within the Barbican database. Software based crypto plugins offer a scalable solution, but are vulnerable to system software attacks. Hardware Security Module (HSM) [9] plugins offer strong security guarantees, but they are expensive and don't scale well. We propose to build an Intel® Software Guard Extension (Intel® SGX) based software crypto plugin that offers security similar to an HSM with the low cost and scalability of a software based solution. We extend OpenStack Barbican API to support attestation of an Intel® SGX crypto plugin, to allow clients higher confidence in the software they are using for storing keys. In addition, the API provides support for mutual attestation for Intel® SGX enabled clients, multi-user key distribution, and extensions for protecting the confidentiality and integrity of the backend database.

## 1. INTRODUCTION

OpenStack [2] is becoming a cloud OS of choice in cloud space. To provide strong data protection in the cloud, encryption is typically used to protect the data in transit as well as data at rest. Key management continues to be a big challenge for all the cloud components that use encryption. In the OpenStack environment, there is a shift towards using a centralized key management system called Barbican [1].

Intel® SGX [8] is a new technology that allows applications to create areas in the application address space to run code where integrity and confidentiality of that code is protected inside the CPU. In this paper we present a solution to build an Intel® SGX based crypto plugin for Barbican. This plugin offers security similar to an HSM with the low cost and scalability of a software based solution. In addition, we also extend OpenStack Barbican API to support attestation of SGX crypto plugin, to allow clients higher confidence in the software they are using for storing keys. Optionally, for Intel® SGX aware clients, we also support mutual attestation, where Barbican plugin can attest to the client before releasing the secret. Our extended REST API also provides supports for multi-user key distribution, where the owner of the secret can specify additional parties that are allowed to extract secrets, based on a well-defined access control list.

We seamlessly integrated our solution into the OpenStack environment and all components using Barbican can now be configured to use an Intel® SGX based key store with no impact to the rest of the system.

Our contributions in the paper are:
- Architecture and design of a new Intel® SGX based crypto plugin for OpenStack Barbican called BarbiE.
- Complete end-to-end solution with support for attestation of BarbiE as well as mutual attestation for Intel® SGX enabled clients.
- Support for multi-user key distribution to achieve secure key distribution among third-party trusted enclaves irrespective of their owners.
- Protection of data with confidentiality and integrity while on storage on a backend database.
- Evaluation of BarbiE (the Barbican Enclave) in a scalable cloud deployment of OpenStack Mitaka [10].

## 2. Background

### 2.1 OpenStack Overview

OpenStack is an open-source cloud operating system designed for public and private cloud infrastructure deployment and management that enables the use of multivendor commodity hardware and relies on software to provide an "eventually consistent" means of fault tolerance/high availability. It consists of many subprojects, each targeted at managing independent aspects of the cloud OS.

The OpenStack platform at its heart consist of 6 core projects that provide various services. Core projects are those that implement OS functionality that meets both public and private cloud deployment needs and is deemed the single best implementation by the OpenStack community. Each of these projects can be instantiated and run separately as they have been designed to be plug-and-playable, allowing users to pick and choose components to meet their needs. However, these projects really work best together. Let's quickly examine the current state of the OpenStack platform core.

The *Nova* project abstracts computing resources in an immensely scalable fashion and is able to run on physical hosts from a multitude of vendors while exposing compute resources via a multitude of virtualization vendors (e.g. VMWare, XEN, KVM (Kernel Virtual Machine), etc.). *Swift* provides scalable and redundant object storage purely implemented in software allowing for low cost commodity drives to be used while maintaining high availability. Networking resources are abstracted by the *Neutron* project which exposes software defined networks to address scalability of deployments. *Cinder* exposes block storage capable of being backed by different types of storage technologies (supports standard Linux server, EMC, Hitachi Data Systems and IBM Storage). *Glance* is the subsystem for managing disk images in OpenStack allowing for a deployment to manage itself with minimal downtime. Refer to the OpenStack documentation [13] to learn about other optional projects as there are a handful not listed above.

Authentication and access control is managed by the *Keystone* project [14]. Keystone hosts a central user directory that is mapped to a services registry that allows access based querying of resources. Keystone is the common authentication system for OpenStack but can also be integrated with other backend authorization systems. Before a user can interact with any of the OpenStack projects, it needs to get a token from Keystone and then that token is used for authentication for interacting with various OpenStack components, including Barbican.

## 2.2 Intel Software Guard Extensions

Intel® SGX is a new processor technology that allows applications to create secure areas in memory called enclaves. These enclaves provide applications with a trusted execution environment where confidentially and integrity of any code and data inside the enclave is protected inside the CPU package boundary. In addition, Intel® SGX also provides a mechanism to remote parties to verify the identity of the enclaves via a process called attestation.

### 2.2.1 Overview

Intel® SGX provides a trusted execution environment for security sensitive applications. It provides applications the ability to create areas in memory called enclaves that provide confidentiality and integrity for the code/data running inside it, even in the presence of buggy or malicious privileged software. An application developer partitions the code into trusted and un-trusted parts. The un-trusted part of the application loads the code that needs trusted execution environment inside an enclave. Once the enclave is built and initialized, no-one can tamper with the code and data inside it with access protected by the CPU.

*2.2.2 Attestation*

Attestation is a process that allows remote parties to verify the identity of a piece of software. Intel® SGX supports attestation [11] where any code running in an enclave can generate a report, which includes the enclave's identity and, with the help of an Intel® SGX SDK provided architectural enclave called a Quoting Enclave, can generate attestations, called quotes, to send to a remote party for verification. The remote party can use Intel's SGX attestation hosted service to verify the quote before provisioning any secrets into the enclave.

*2.2.3 Sealing*

Intel® SGX supports a mechanism called sealing [12] for cryptographically binding any secrets to an enclave. Any enclave can use a hardware generated key, called a seal key, to encrypt the secrets with the key. The key is only available to the owner enclave running on the same platform.

## 2.3 OpenStack Barbican

*2.3.1 Overview*

OpenStack Barbican [3] is a platform developed by the OpenStack community aimed at providing cryptographic functions useful for all environments including large ephemeral clouds. Barbican exposes REST APIs designed for the secure storage, provisioning and management of secrets such as passwords, encryption keys and X.509 Certificates [5].

*2.3.2 Architecture*

Figure 1 below describes the high-level architecture of OpenStack Barbican (larger views of all figures in this paper are in the appendix section)

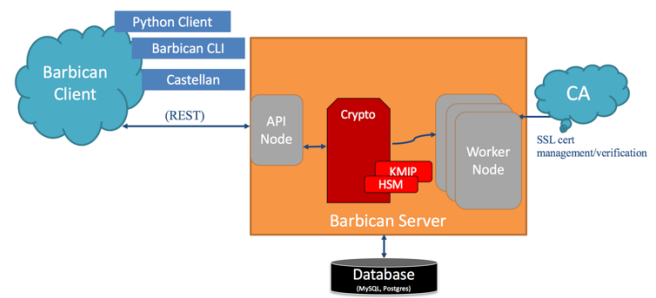

**Figure 1. OpenStack Barbican**

Clients interact with API nodes using REST APIs. API nodes can directly service some of the requests synchronously, otherwise they put the requests in messages queues to be handled by the worker queues asynchronously. This architecture supports scalable design where multiple API and worker nodes can be added/deleted as needed.

### 2.3.2.1 Barbican storage mode for Secrets (plugins)

Barbican supports two modes for storing secrets. In secret mode, Barbican offloads both encryption/decryption and encrypted secure storage to a plugin and is not responsible for storing any encrypted secrets in its own data store. Some examples include Redhat Dogtag service and Key Management Interoperability Protocol (KMIP) [15]. In cryptographic mode, Barbican supports in-memory crypto or an appliance like an HSM to perform encryption/decryption. In this mode, Barbican core stores encrypted secrets in Barbican's data store. Some examples include Barbican in-memory simple-crypto plugin or a PKCS11 based interface to an HSM.

### 2.3.2.2 Barbican Secret Management

Cloud users can offload sensitive parts of their application by securely storing them in Barbican. Barbican further protects these user secrets in a backend database using encryption keys. Barbican can also generate sensitive information, from AES keys for disk encryption to X.509 certificates used to secure application servers and load balancers. To bootstrap the server's ability to securely manage user secrets, an administrator must provision a Key Encryption Key (KEK) per project. The KEK is used to encrypt/decrypt secrets as they are placed into and retrieved from the backend DB. In addition, session keys (SK) are used to protect any data in flight after a secure session has been established.

## 3. Intel® SGX Enabled KMS Architecture

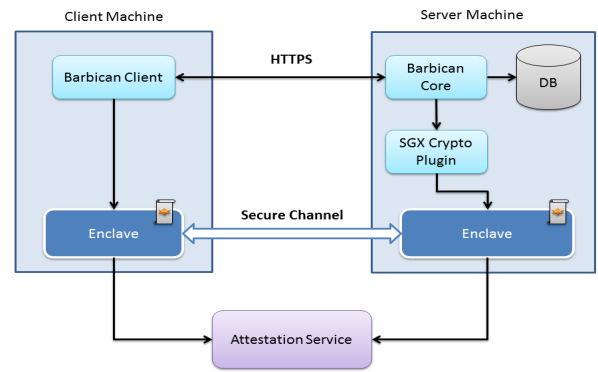

**Figure 2. High Level Architecture**

As shown in Figure 2, Intel® SGX enabled KMS system is architected to provide complete protection of secrets while at rest, in use, and in transit. As discussed in section 2.2, Intel® SGX provides hardware root of trust capabilities that provide integrity and confidentiality of data inside an enclave as well as allow an enclave to prove its trustworthiness to a challenger.

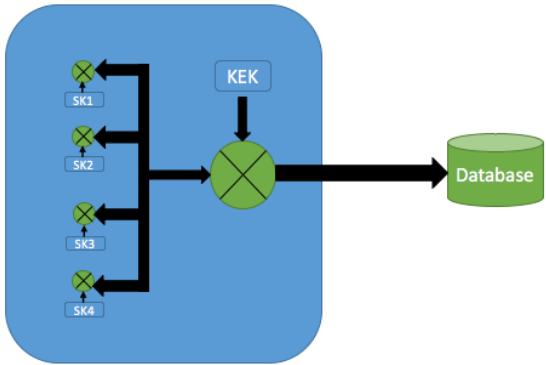

**Figure 3. KEK Provisioning**

The SGX Barbican server (a.k.a, Intel® SGX KMS, used interchangeably) implements an Intel® SGX enclave that is responsible for performing all cryptographic operations, manage encryption keys, and seal objects into databases for secure storage. This enclave is also responsible for proving its trustworthiness to external applications that intend to store application secrets and encryption keys in the Intel® SGX KMS server. Once the client establishes trust with the Intel® SGX KMS system using the Intel® SGX remote attestation protocol, it can also request the Intel® SGX KMS system to generate PKI and symmetric keys on the application's behalf.

Since OpenStack Barbican implements a plugin model for various modes of operation, the Intel® SGX enclave is wrapped as a crypto plugin that fits inside the Barbican core stack. The Intel® SGX enclave itself is built as a C/C++ library and the plugin is implemented in python with python to C CFFI bridging between the two. The Intel® SGX crypto plugin layer hides and abstracts Intel® SGX specific details from the Barbican core system. As shown in figure 3 Intel® SGX enclave is responsible for securely storing client secrets and it uses a master key called KEK for that purpose. The KEK is either generated inside the enclave during initialization or provisioned remotely by the administrator for a scale out deployment. The KEK is kept sealed using enclave's sealing feature, as opposed to plain text format of simple crypto plugin in legacy Barbican.

The database is also enhanced to enforce a strict to liberal level of access to secrets under a particular tenant/project. This is done by using the Intel® SGX security properties like signer, measurement, ISV_SVN, and CPU_SVN as identity parameters for access authentication. In addition, for non Intel® SGX clients, keystone token ID and client public key could also be used for identification and access control.

However, to use the Intel® SGX KMS in a truly secure manner, we recommend that the client application also use an enclave on its side to interact with the server. This would allow the Intel® SGX KMS to be able to securely authenticate the client using the Intel® SGX remote attestation process. However, to maintain compatibility with legacy clients, the Intel® SGX KMS server seamlessly supports legacy APIs with no change in the message exchange format.

## 3.1 Barbican Clients

Our goal for cloud infrastructure providers is to make the transition from a legacy Barbican system to an Intel® SGX enabled Barbican system as smooth as possible with minimal or no impact to their current customer base. Therefore, we consider the following types of clients to be interacting with the Intel® SGX KMS system.

### 3.1.1 Legacy Barbican Clients

These are the Barbican clients that are not aware of Intel® SGX but still be able to talk to the SGX Barbican server without any changes using the current v1 (version 1) interface (e.g., GET /v1/secrets). These clients will still use the legacy protocol to talk to Barbican and perform key management. The SGX Barbican server will encrypt the secrets with a KEK inside the enclave. The REST interfaces remain the same for these clients and the enclave in the SGX Barbican server can be updated without notifying these clients.

### 3.1.2 Intel® SGX Aware Barbican Clients

These are new Barbican Clients that are not Intel® SGX enabled but rely on the presence of the Intel® SGX enclave on the SGX Barbican server and challenge the server through a remote attestation protocol before storing any secrets. The attestation APIs are available as standard REST v2 APIs (ex. GET /v2/secrets) and the response from the server, which would be an Intel® SGX quote, can be validated using Intel Attestation Service. These clients also support multi-user key distribution by setting an access control list while storing the secrets.

### 3.1.3 Intel® SGX Enabled Barbican Clients

These are new Barbican Clients that are also Intel® SGX enabled. These clients support all the features supported by Intel® SGX aware clients, but they also require the SGX Barbican server to challenge the client before releasing the secrets back to the client.

## 3.2 SGX Barbican Server Crypto Plugin (BarbiE)

The Intel® SGX hardened Barbican crypto plugin was designed to enhance the protection of user secrets while in use, at rest, and in transit to/from the client/server. To achieve these objectives, the key Intel® SGX features that are employed are memory protection, sealing, and remote attestation. Intel® SGX sealing is a feature in which the hardware is able to generate a key for an enclave that is tied to a unique enclave identity or signature. This allows for an enclave to keep secrets by using this key to encrypt data it wishes to be protected when outside of the enclave boundary (e.g., in main memory or at rest on the disk). By leveraging Intel® SGX remote attestation, we are able to securely provision a key from the Barbican client to the Intel® SGX Barbican server. In our case, this key will be the secure channel key (SK), which will be used to encrypt/decrypt traffic in flight.

At the heart of our hardened Barbican implementation is an Intel® SGX enclave, which we have named BarbiE (the Barbican Enclave). Per the Intel® SGX programming model, there is an untrusted application responsible for hosting or loading the trusted enclave within its virtual address space. The Intel® SGX enabled Barbican crypto plugin exercises the trusted actions provided by BarbiE by communicating with the untrusted part of the SGX Barbican crypto plugin application via CFFI (C Foreign Function Interface). These actions exposed by the untrusted part of the plugin can be either untrusted or trusted. In the case of a trusted action, the API call will result in an ECALL or an entry into BarbiE to fulfill some task.

### 3.3 Mutual Attestation (MA)

Intel® SGX allows an enclave to prove its trustworthiness to a challenger via remote attestation protocol and BarbiE exposes that support to Intel® SGX aware clients. In addition, we also support BarbiE challenging the receiver of the secret if the client is running inside an enclave on an Intel® SGX enabled client. With enclaves at each end of the transaction we are able to attest to the trustworthiness of both parties involved. We call this method of trust establishment Mutual Attestation (MA).

### 3.4 Multi-User Key Distribution

In a distributed architecture, different components deal with same secrets. The owner of the enclave may want to allow other components in a system to have direct access to the secret. With multi-user key distribution feature, we provide a mechanism for the owner of the secret to provide an access control list to BarbiE for secret distribution. This can be done by using the Intel® SGX security properties like signer, measurement, ISV_SVN, and CPU_SVN as identity parameters for access authentication. In addition, for non Intel® SGX clients, keystone token ID and client public key could also be used for identification and access control.

### 3.5 Protection of backend Database

In OpenStack Barbican, access to secrets by tenants/projects are controlled using ACLs maintained in a Database. These ACLs are present in a tenant ID/project ID mapping table within the Barbican database. The security model in SGX Barbican is that all cryptographic operations are performed only inside BarbiE. Any confidentiality or integrity breach of the database will not directly result in secrets being compromised. However, if an attacker is able to modify the tenant ID/project ID mappings by manipulating the database records, it could grant the attacker unauthorized access to secrets (stored by the Intel® SGX Barbican enclave inside the DB) using Barbican's regular REST APIs. For example, the attacker can alter session keys records in the DB to copy encrypted session key of a user to be attacked in its own row. After successful mutual attestation (as the attacker is a registered user with OpenStack Barbican), the attacker will gain access to session key of user to be attacked. To prevent this from happening, we cryptographically tie the encrypted values in the DB with the tenant/project ID. For this mechanism to work correctly, we also need the tenant/project ID information to be provided by the client securely (through the secure channel). In this way, altering the DB records by an attacker will not work anymore and can be detected. Also, we need a replay protection mechanism in the DB such that any record in the DB when replaced with an old record can also be detected. Replay protection of persistent data has not been implemented in the current version of the prototype, but is supported by Intel® SGX architecture.

## 4. Intel® SGX Enabled KMS Design

In section 2.2 we described Intel® SGX at a high level, Specifically providing a brief introduction of the memory protections, data sealing, and remote attestation (RA). These Intel® SGX features when combined provide protection of secrets at rest, in use, and in transit. Section 3 introduced our architecture specific keys SK and KEK. SK is used to protect secrets in transit to/from the client/server and KEK is used to protect content that is database bound on the server end. This section will explore a potential implementation based on Intel® SGX features, SK, and KEK that allows for secure end-to-end protection of user secrets in a cloud manager such as OpenStack Barbican. We will now describe the implementation of the Intel® SGX Barbican server and its interaction with Intel® SGX aware and Intel® SGX enabled clients.

### 4.1 Session Key (SK) Negotiation

Remote Attestation of an Intel® SGX enclave allows a remote challenger to cryptographically verify the trustworthiness of attesting enclave. In our implementation, BarbiE is an Intel® SGX enclave and clients are able to challenge it via REST APIs as shown in Figure 4.

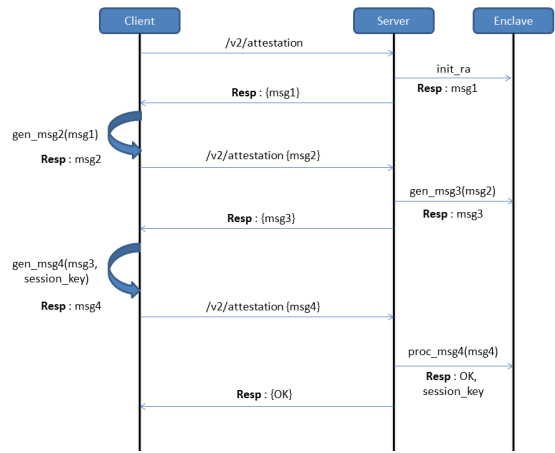

**Figure 4. Remote Attestation REST APIs**

An Intel® SGX aware client can use these REST APIs and the APIs provided by the Intel® SGX SDK to facilitate the RA (MSGX processing/generation) and negotiate an SK. Intel® SGX aware clients should inspect the BarbiE quote/identity prior to completing the RA flow to ensure the BarbiE identity is as expected.

A client running on Intel® SGX capable hardware can instantiate an enclave itself for further protections on the client side. With enclaves at each end of the transaction, we are able to attest to the trustworthiness of both parties involved. We call this method of trust establishment Mutual Attestation (MA).

### 4.1.1 Mutual attestation flow

The flow for negotiating SK during Mutual Attestation is shown below in figure 5.

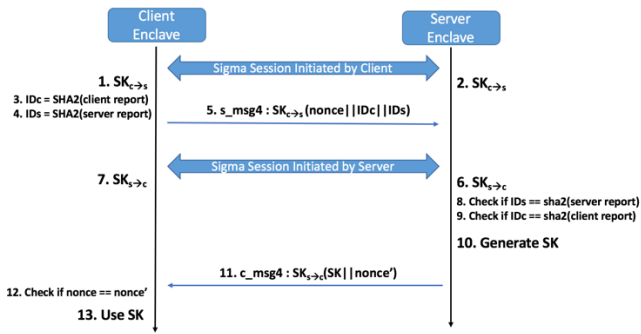

**Figure 5. SK negotiation during Mutual attestation**

During the mutual attestation process, the client enclave first does a remote attestation of the server enclave. During this process both client and server enclaves derive a shared key using the sigma protocol. Client enclave then generates a nonce, calculates SHA2 of its own and server enclave's Intel® SGX report and then sends s_msg4 to the server enclave as shown in Figure 5.

The server enclave now does a remote attestation of the client enclave and verifies the SHA2 signature of the client and its own Intel® SGX reports with the ones sent in s_msg4. If they match, the server enclave is confirmed to be communicating with the correct client enclave. It then generates a session key (SK) and sends both the nonce (received in s_msg4) and the SK in c_msg4 back to the client enclave.

The client enclave does a match of the nonce received in c_msg4 with the one originally sent. If they match, the client enclave is confirmed to be communicating with the correct server enclave and continues to use the SK for secret management.

Details on the sigma session between BarbiE server enclave and client enclave using REST APIs are shown in Figure 6.

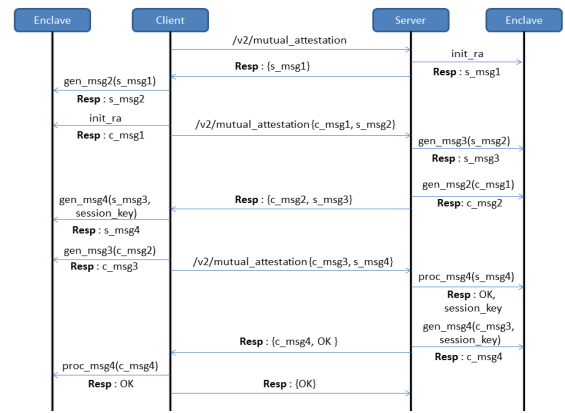

**Figure 6. Mutual Attestation REST APIs**

In case of Mutual Attestation, Barbican server records the client enclave's identity/quote as found in RA MSG3. The quote contains a few additional fields but it only records MRENCLAVE (the measurement of the client's enclave) and MRSIGNER (signer of the client's enclave) fields. These fields will allow the Barbican server enclave to verify and return secrets only to valid Intel® SGX enabled clients. Intel® SGX enabled OpenStack Barbican is enhanced to support three types of policy enforcement. This is done by adding new model in database with SK, policy, MRENCLAVE and MRSIGNER for corresponding tenant/project. Policy can be set after a successful RA or MA. RA supports policy number 3 and MA supports policies 1, 2 and 3. Policy type combined with MRENCLAVE and MRSIGNER determines the level (strict to liberal) of authentication and authorization.

Policy 1 : Match measurement of the client enclave. If a project policy is set to '1' then MRENCLAVE of client enclave must match with the MRENCLAVE of the project owner stored in database when the policy was set.

Policy 2 : Match signer of the client enclave. If a project policy is set to '2' then MRSIGNER of client enclave must match the MRSIGNER of the project owner stored in database when the policy was set.

Policy 3 : Match multiple client enclave measurements. If a project policy is set to '3' then MRENCLAVE of client enclave must match with one of child MRENCLAVEs stored by the project owner in database when the policy was set.

In either case ISV_SVN should also be taken into consideration such that the retriever ISV_SVN is >= that of the storing party's ISV_SVN to prevent compromised enclaves from accessing the secrets of patched enclaves.

### 4.2 Multi-User Key distribution

In a distributed architecture, different components deal with same secrets. In OpenStack for example, Cinder block storage might be the entity that generates/stores a secret in Barbican for encrypted volumes, but Nova compute node is the entity which actually uses that secret and needs access to it. In this scenario, Cinder and Nova clients will have different MRENCLAVEs. Hence, there needs to be a mechanism to share a session key(SK) with multiple enclaves under same tenant/project.

This is achieved by policy '3' enforcement in Intel® SGX enabled OpenStack Barbican. This policy verifies the MRENCLAVE of the requesting client (including the identity of the owner enclave passed by the client) with the list of MRENCLAVES stored in the database for the tenant/project by the owner. This list specifies the enclaves to which the secrets will be accessible.

The access control design can also be easily extended to achieve per secret access control policy rather than per project/tenant policy.

## 4.3 Secret Management

All secrets sent to the Barbican server via an Intel® SGX aware client require an SK (session key) be negotiated via either RA (remote attestation) or MA (mutual attestation) to protect secrets in flight. Let's now explore how the SK is used to provision the KEK (key encryption key, or master key) and how both SK and KEK are used in secret store and retrieval flows. The KEK is the key material used for all cipher text in/out from the backend DB. No plain text is stored in the DB and the KEK is protected in flight via the SK and at rest on the server disk using sealing. Figure 7 shows the REST interactions exposed to an admin client to facilitate the KEK provisioning. The admin simply needs to perform an RA or MA to negotiate an SK, then provision the SK encrypted KEK (SK_KEK) to the server, at which point the server stores the KEK sealed on disk such that it is recoverable across restarts of the enclave without admin intervention. This KEK can also become the master KEK for the entire setup. In that case the individual projects/tenants would get their own individual KEKs generated and sealed on disk using the master KEK

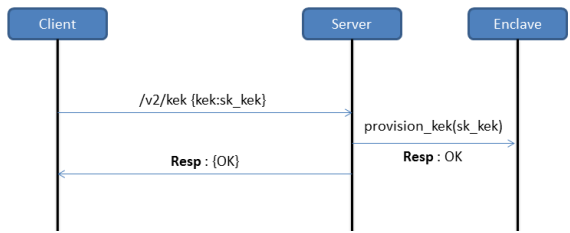

**Figure 7. Master Key (KEK) Provisioning REST APIs**

Once a KEK is provisioned, the server is ready to process secret management requests (store/retrieval) on behalf of its users. Figure 8 shows the secret management flow, all of the required artifacts, and the cipher text transformations, while Figure 9 depicts the API flows.

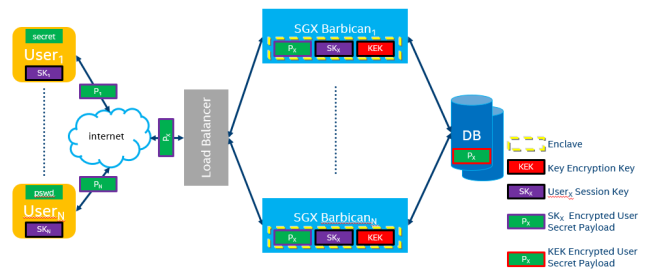

**Figure 8. High Level Architecture**

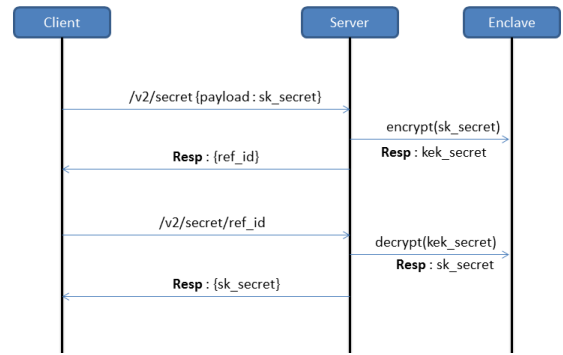

**Figure 9. Secret Management REST APIs**

The client and server must first negotiate an SK via either RA or MA. The client wishing to store a secret should then encrypt the said secret (generating SK_SECRET) with the SK and then via the /v2/secret REST API send the cipher text and metadata to the server. The server will (within BarbiE) decrypt the user secret with the SK and encrypt with KEK (generating KEK_SECRET) which will produce cipher text that is DB appropriate and should be stored along with the quote retrieved during SK negotiation in the case of MA. Upon successful record insertion a reference ID (REF_ID) is returned as the storage request response. A client wishing to retrieve a secret presents a valid REF_ID to the server. In case of an RA provisioned secret, the server refers to the policy set for allowing access to SK_SECRET. Refer to section 4.1 for details. However, the owner entity of the secret can still access the secret using the original session key negotiated during the RA session. In case of a secret provisioned via MA, the server will challenge the client enclave via reverse RA (see section 4.1.1) to obtain the SGX parameters of the caller. At this point a policy based decision will be made regarding whether or not to return the SK_SECRET to the client. As previously stated in section 4.1, the choice of fields in the quote used for verification is implementation specific.

### 4.4 Database Integrity protection

To bind session keys encrypted with a master key with its respective tenant/project we use AES GCM Additional Authentication Data (AAD) mechanism. During Mutual Attestation, client needs to provide its tenant/project ID as part of msg4, which is to be sent to the server enclave. This will help in securely provisioning the ID from client to server. This ID will then be used as AAD during encryption of tenant/project specific session key inside enclave before storing them to the DB. As tenant/project ID is part of Mutual Attestation, for an attacker to have successful MA, it would have to provide its own ID in msg4. Even with this, they won't be able to gain access to session keys of other users even after altering the DB due to ID mismatch during decryption.

### 5. BarbiE Cloud Deployment and Usage

A single cloud service implementation has a finite capacity, which can lead to runtime exceptions and performance degradation when its processing thresholds in terms of compute power and available memory are exceeded.

There are two ways to overcome this problem. One method would be scale-up the deployment that involves adding more compute power, memory and I/O bandwidth to the same physical instance. The other way is to scale-out by running redundant deployments of the same instance. Typically, in this type of deployment, a load balancing system is added to dynamically distribute workloads across cloud service implementations.

The multiple cloud service implementations are organized into a resource pool. As shown in figure 10 the load balancer is configured as an external or consumer facing component, allowing internal service applications to balance workloads among themselves.

Depending on the type of application, the load balancer can be configured stateless or stateful wherein state information inside

HTTP headers can be used by the load balancer to make more intelligent balancing decisions. The load balancer gets all the messages sent by cloud service users and forwards the messages at runtime to the actual service instances in a way such that the workload processing is horizontally scaled.

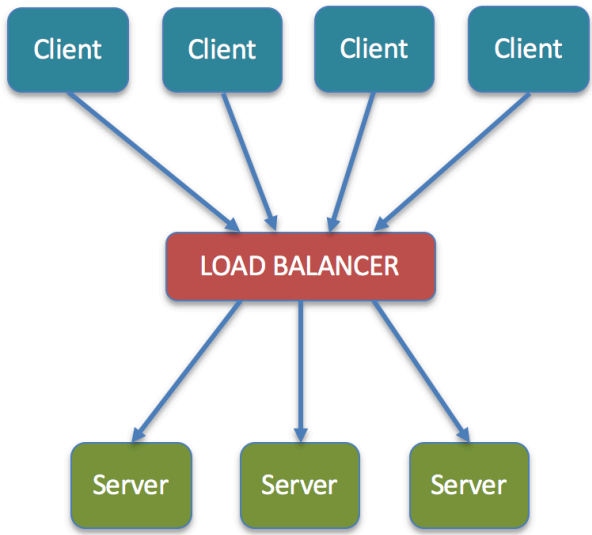

**Figure 10. Cloud Deployment**

A point to note here is that a load balancer can be positioned either independent of the cloud service or be built-in as part of the application. However, when built-in as part of the application, the service applications will be responsible to form a cluster of master and slave nodes where the master node takes the decision of routing requests to slave nodes based on the request.

An SGX Barbican server would also be predominantly deployed as a cloud key store to allow cloud users to be able to offload confidential or sensitive aspects of their projects to a secure storage.

Cloud based applications can replace sensitive information such as encryption keys, database passwords, and unique hyperlinks, etc., that are securely stored in SGX Barbican for later retrieval. SGX Barbican encrypts that sensitive information inside an Intel® SGX enclave, which is then stored inside a DB. The encryption key used by the Intel® SGX enclave (aka KEK) is provisioned by the administrator using the remote attestation protocol [11]

Alternatively, for a single instance SGX Barbican server, the KEK provisioning step can be made optional as the Intel® SGX sealing key (or a key derived from the sealing key) could be used as the KEK. However, once these Intel® SGX Barbican server instances are scaled out on different hardware platforms, the sealing key KEK method will not work anymore as Intel® SGX enclaves running on different hardware platforms will derive different KEKs. Hence, on a scaled design, it is the responsibility for the administrator to provision the same KEK in all Intel® SGX Barbican instances before bringing them online. This way, all Barbican instances are identical and stateless accessing a common database and any of the instances can handle requests in any order as all transaction states are maintained in the database.

Figure 12 shows how a typical scaled SGX Barbican setup would look like

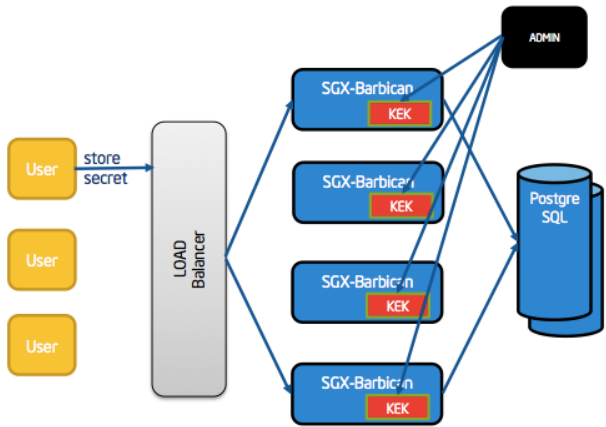

**Figure 11. SGX Barbican Scaled Deployment**

The following steps show how this setup is initialized and used by a user or a cloud application

**Step 1.** SGX Barbican instances are created (inside VMs or containers)

**Step 2.** All those instances point to the same database system.

- DB system can also be distributed for high availability and throughput

**Step 3.** Administrator performs remote attestation with each and every SGX Barbican enclave instance to gain trust of the enclave and establish a secure encrypted channel for data transfer.

**Step 4.** Admin uses the admin interface to provision KEK on all SGX Barbican instances using the secure channel established in the previous step

**Step 5.** An HTTP load balancer is configured to connect to all these instances on an internal network.

**Step 6.** User connects to load balancer to perform remote attestation of SGX-Barbican

**Step 7.** Load balancer chooses a routing policy (random, load based or round robin) to route the request to any instance as they are identical and store internal states in a common DB.

**Step 8.** User completes remote attestation with instance x and establishes a secure channel. Both parties now have a shared key (SK). Optionally, user can also set multi-user access control based on SGX measurement of third party enclaves. Refer to 4.1 and 4.2 for more details

**Step 9.** Instance x encrypts shared key using the KEK and stores it in DB

**Step 10.** User now sends request for "secret store" by encrypting the user payload with the shared key

**Step 11.** Request goes to instance#1

**Step 12.** Instance#1 finds shared key encrypted in DB

**Step 13.** Instance#1 decrypts user payload using shared key and encrypts it with KEK.

**Step 14.** Instance#1 stores the encrypted user secret in DB

**Step 15.** User now requests previously stored secret from SGX-Barbican system. Third party access to the secret is allowed by SGX enabled clients based on the access control policy set on the project (refer section 4.1 and 4.2)

**Step 16.** Request goes to a different instance this time. E.g., instance#4

**Step 17.** Based on the request, instance#4 either does an attestation of the user enclave or decrypts shared key and user secret with KEK. If the request is for attestation, then it sends the shared key over the secure channel after a successful policy matching attestation flow

**Step 18.** Instance#4 encrypts user secret with shared key and returns the payload to the user

**Step 19.** User decrypts payload with shared key to retrieve original plain text secret.

Intel's SGX remote attestation protocol requires a quoting enclave service that signs the report of application enclaves to generate a quote. This quote contains information to establish a shared key using Diffie-Hellman(DH) method [16]. The DH protocol requires multiple message exchanges between the remote party and enclave endpoint. The following figure shows a high level flow of SGX remote attestation protocol between challenger and application enclave.

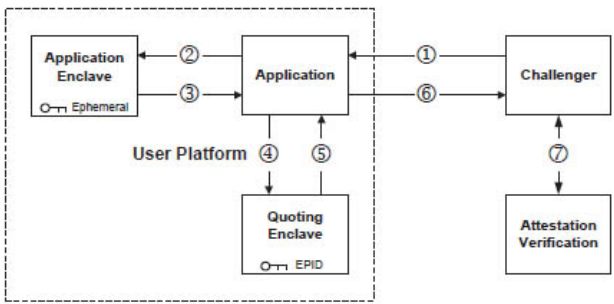

**Figure 12. SGX Remote Attestation**

SGX Barbican system implements these message exchanges as REST APIs. The REST APIs for remote attestation are themselves stateless, but carry context references in the message body to complete the remote attestation protocol. The following time-line flow diagram shows how individual components of the SGX Barbican system interact with each other using REST APIs to establish shared key based secure channel [11].

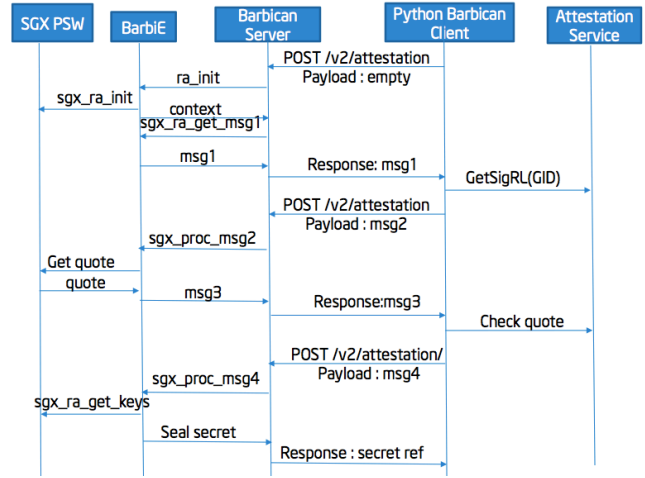

**Figure 13. SGX Remote attestation on scaled deployments**

On a load-balanced setup, however, sequential requests cannot be guaranteed to land on the same instance. In fact, the load balancer ensures that sequential requests are distributed across different service instances to balance the load. Therefore, remote attestation REST calls pertaining to the same session cannot be guaranteed to land on the same instance and hence are bound to fail.

To work around this issue, we must configure some stickiness (e.g., a cookie) in the load balancer that is sent back during the first response. The cookie basically tells the load balancer which node to select for routing the request. During the remote attestation process, the client ensures to use the same cookie for all remote attestation calls made to establish a secure channel. Once a secure channel has been established, the client can ignore the cookie for all subsequent transactions.

## 6. Threat Model

KEK in Barbican is used to protect user secrets and we make sure that we provide confidentiality and integrity guarantees for KEK. In addition, session key (SK) is used to protect any data in flight after a secure session has been established. In the following sections, we will discuss the threats and mitigations against various assets managed by SGX enabled Barbican.

### 6.1.1 Confidentiality Threats and Mitigations

| Asset | Attack | Mitigation |
|---|---|---|
| KEK | Adversary accesses KEK | On Server: Protected<br>-In use by the enclave<br>-At rest via sealing<br>-In flight via RA DH<br>On Admin machine: Out of scope |
| SK | Adversary accesses SK | On Server and SGX enabled Client: Protected<br>-In use by the enclave<br>-At rest via KEK encryption<br>-in flight via RA/MA DH<br>On Legacy Client: Out of Scope |
| User Secrets | Adversary accesses user secrets | On Server and SGX enabled Client: Protected<br>-In use by the enclave<br>-At rest via KEK encryption<br>-in flight via SK encryption<br>On Legacy Client: Out of Scope |

**Table 1. Confidentiality Assets, threats & mitigation**

### 6.1.2 Integrity Threats and Mitigations

| Asset | Attack | Mitigation |
|---|---|---|
| KEK | Adversary manipulates KEK | On Server: Protected<br>-In use by the enclave<br>-At rest via sealing – subject to replay attack<br>-In flight via RA DH<br>On Admin Machine: Out of scope |
| SK | Adversary manipulates SK | On Server and SGX enabled Client: Protected<br>-In use by the enclave<br>-At rest via KEK encryption – subject to replay attacks<br>-In flight via RA/MA DH<br>On Legacy Client: Out of Scope |
| User Secrets | Adversary manipulates user secrets | On Server and SGX enabled Client: Protected<br>-In use by the enclave<br>-At rest via KEK encryption<br>-In flight via SK encryption<br>On Legacy Client: Out of Scope |

## Table 2. Integrity Assets, threats & mitigation

### 6.1.3 Authentication Threats and Mitigations

| Asset | Attack | Mitigation |
|---|---|---|
| Legacy Client Identity (Authorization Token) | Adversary accesses Authorization token | On Legacy Client and Server: Out of Scope (Legacy Authorization agent protections) |
| Client Enclave Identity | Adversary manipulates enclave identity | On Legacy Client: N/A<br>On SGX enabled Client: Protected<br>On Server:<br>-In use by the enclave<br>-At rest via KEK encryption – subject to replay attacks<br>-In flight via MA DH |
| Server Enclave Identity | Adversary manipulates enclave identity | On Legacy Client: Out of Scope<br>On SGX enabled Client: Protected<br>-In use by the enclave<br>-At rest via sealing<br>-In flight via RA/MA DH<br>On Server: SGX protections |
| SK | Adversary accesses SK | On Legacy Client: Out of Scope<br>On Server and SGX enabled Client: Protected<br>-In use by the enclave<br>-At rest via KEK encryption<br>-In flight via SK encryption |

**Table 3. Authentication Assets, threats & mitigation**

### 6.1.4 Database Threat and Mitigation

| Asset | Attack | Mitigation |
|---|---|---|
| Legacy Client Identity (Authorization Token) | Not stored in DB | NA |
| Client Enclave Identity | Adversary manipulates DB records swapping/replacing with attacker's enclave Identity | On Legacy Client: N/A<br>On SGX enabled Client: Protected<br>On Server:<br>-In use by the enclave<br>-In DB via AAD with projectID<br>-in flight via MA DH |
| Server Enclave Identity | Not Stored in DB | NA |
| SK | Adversary manipulates DB records swapping/replacing own encrypted SK with required encrypted SK | On Legacy Client: N/A<br>On SGX enabled Client: Protected<br>On Server:<br>-In use by the enclave<br>-In DB via AAD with projectID<br>-In flight via MA DH |
| KEK | Not Stored in DB | NA |

## 7. Performance Evaluation

We performed Apache benchmark testing on a standalone as well as a scaled setup under various load conditions. Table 4 captures the benchmark numbers for single Xeon E3 3.7Ghz node running legacy Barbican and SGX Barbican. The scaled setup was tested with 4 VMs on the same platform

| Metric | Legacy Barbican | SGX Barbican | 4x SGX Barbican |
|---|---|---|---|
| Concurrency Level | 2 | 2 | 2 |
| # of requests per user | 100 | 100 | 100 |
| # of users | 5 | 5 | 5 |
| Mean Processing Time | 339ms | 376ms | 101ms |
| Mean Connect time | 3ms | 3ms | 5ms |
| Mean time per request | 343ms | 378ms | 107ms |
| # Req. processed per second | 5.83 | 5.28 | 18.68 |
| Mean time across all connections | 171ms | 189ms | 53ms |
| Total body sent | 22100 bytes | 27900 bytes | 27900 bytes |
| Total time taken to complete the test | 17.1 seconds | 18.9 seconds | 5.3 seconds |

Table 4. Single node performance

For the 4x scaled SGX Barbican setup, we also performed high request count tests; the Apache benchmark numbers for the tests are listed in Table 5.

| Metric | 4x SGX Barbican |
|---|---|
| # of users | 25 |
| Concurrency Level | 5 |
| Requests per user | 500 |
| # Req. processed per Sec | 2.69 |
| Time taken per request | 1859ms |

Table 5. High SGX Barbican high request count performance

## 8. Use Cases

An SGX Barbican server can be used for any existing applications that use Barbican for their secret management, without any modifications, and will a get higher level of protection from attacks that a non-SGX Barbican server. With modifications, these applications will benefit using mutual attestation and multi-user key distribution feature to build secure end-to-end solutions. In addition, an SGX Barbican server can also be used independently of OpenStack as a standalone Intel® SGX protected KMS, and can be used for any secret management in the cloud. We will highlight some sample usages in the rest of this section.

### 8.1 OpenStack Cinder

Cinder is block storage service for OpenStack. Cinder uses Barbican for managing keys for encrypted volumes and communicates with Barbican during encrypted volume creation. The volume encryption key is never exposed to Cinder, but is managed by Barbican. While attaching a volume to an instance, the key is release to Nova, where it is in the clear during use. We are exploring an Intel® SGX enabled file system (based on EncFS) to protect the key even while in use on the Nova compute node. Similar usage can also be applied to OpenStack Swift for object store protection.

## 8.2 OpenStack Neutron

Neutron is yet another OpenStack project to provide "networking as a service" between interface devices. Neutron uses Barbican for protecting TLS/SSL private keys for TLS/SSL termination load balancer. SGX Barbican can provide much higher guarantees for protection of such private keys.

## 8.3 Secure Message Queues

A message queue is an asynchronous communication mechanism between different applications in the cloud. For applications that are security sensitive, we can support Intel® SGX enabled secure message queues; messages are always stored in encrypted format and only valid consumers get access to it.

The key management of such message queues can be handled by SGX Barbican. Intel® SGX aware message queue publishers will perform mutual attestation with SGX Barbican server to get a SK. The SGX Barbican server then encrypts messages inside its enclave with the SK and publish them in queue for consumers. The publisher can also set policy in the SGX Barbican server to allow messages to be accessed only by trusted subscriber enclaves. The subscriber enclaves will also perform mutual attestation with the SGX Barbican server. After a successful mutual attestation, the SK will be shared based on the ACL. Messages are then de-queued and decrypted inside subscriber enclaves.

## 8.4 Secure Docker Secrets

Docker supports protection of secrets such as password, keys, and certificates in Docker Secrets. Docket Secrets can be enhanced directly to use Intel® SGX to gain protection of secrets on disk as well as in use. Another option is to encrypt secrets that are stored in Docker Secrets with a master key stored in an SGX Barbican server and then using the multi-user key distribution feature to allow Intel® SGX enabled consumers of the Docker Secret to retrieve the master key, so as to be able to decrypt the secret for use. This will allow end-to-end protection of Docker Secrets with minimal modifications to Docker's infrastructure.

## 8.5 Secure DRM key management

Encryption is typically used to protect media playback data. Key management continues to be a big problem in this space. SGX Barbican can be used on the edge appliances where keys can be provisioned by the backend and all the media is encrypted using keys stored and protected by SGX Barbican.

## 8.6 Secure Telco workloads

Many Telco workloads that are moving to the cloud use encryption for data protection in OpenStack environments. We envision many usages in Telco space that can benefit from our Intel® SGX enabled Barbican KMS.

## 9. Related Work

Fortanix Self-Defending Key Management Service (SDKMS) [6] claims to be the first cloud service secured with Intel® SGX. With SDKMS, you can securely generate, store, and use cryptographic keys and certificates, as well as secrets, such as passwords, API keys, tokens, or any blob of data. Applications and containers can integrate with SDKMS using legacy cryptographic interfaces or using its native RESTful interface. It's not very clear from their documentation if they offer Mutual attestation, multi-user key distribution or integrity protection of backend database.

## 10. Conclusions and Ongoing Work

In this paper we demonstrated the feasibility of a scalable and performant cloud-based KMS that offers almost the same security guarantees as a hardware HSM, along with the scalability and low cost of a software-based solution. We demonstrated support for mutual attestation as well as multi-user key distribution, allowing enhanced authentication as well as ease of secret distribution in a distributed cloud environment. In addition, we implemented support for integrity protection of the backend database.

As a next step, we plan to evaluate Intel® SGX support in other components of OpenStack to allow end-to-end protection of secrets. We are also evaluating use of SGX KMS as a standalone KMS for use with other systems that require key protection.

## 11. ACKNOWLEDGMENTS

Many thanks to Michael Steiner, Nilesh Somani, Shweta Shinde and Sudha Krishnakumar and many other reviewers for their valuable feedback on this paper.

# Appendix

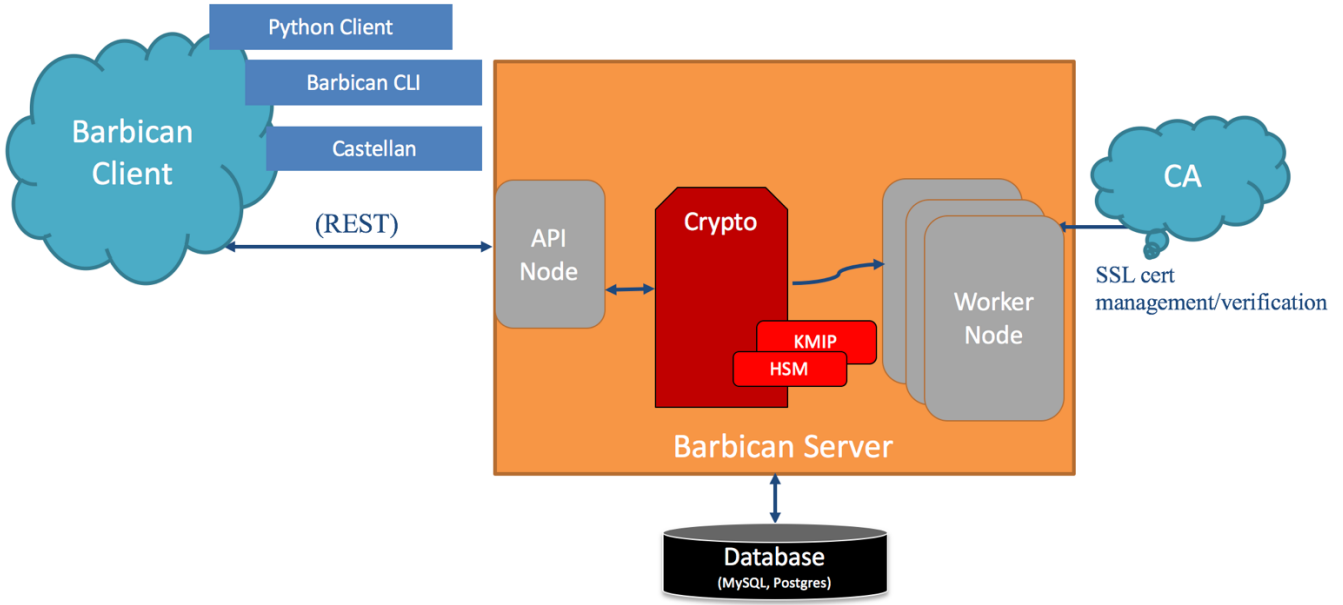

**Figure 1. OpenStack Barbican**

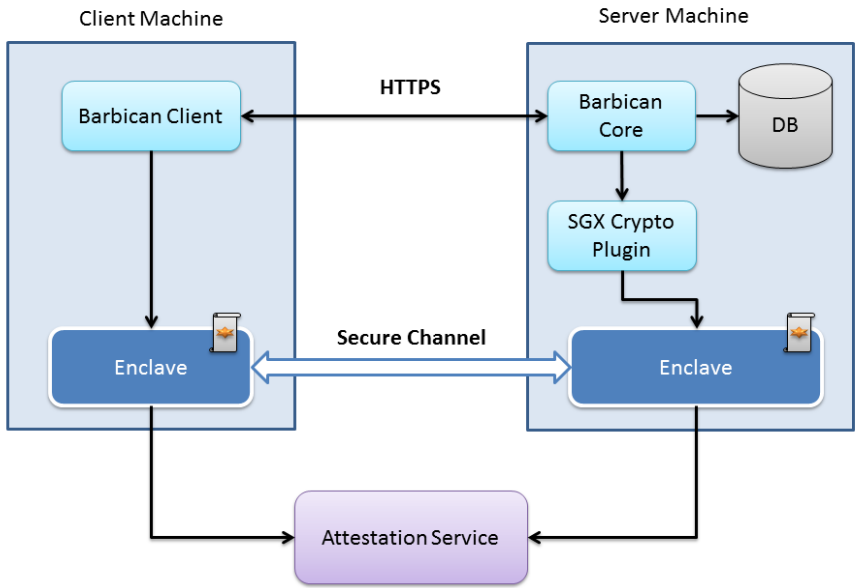

**Figure 2. High Level Architecture**

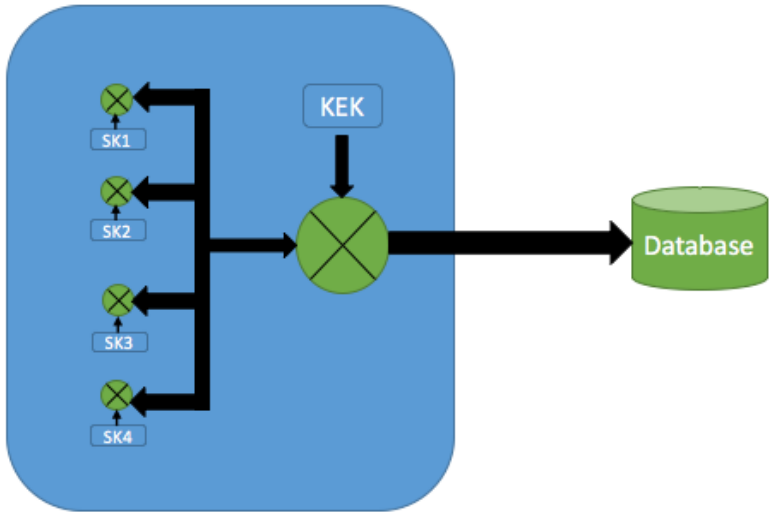

**Figure 3. KEK Provisioning**

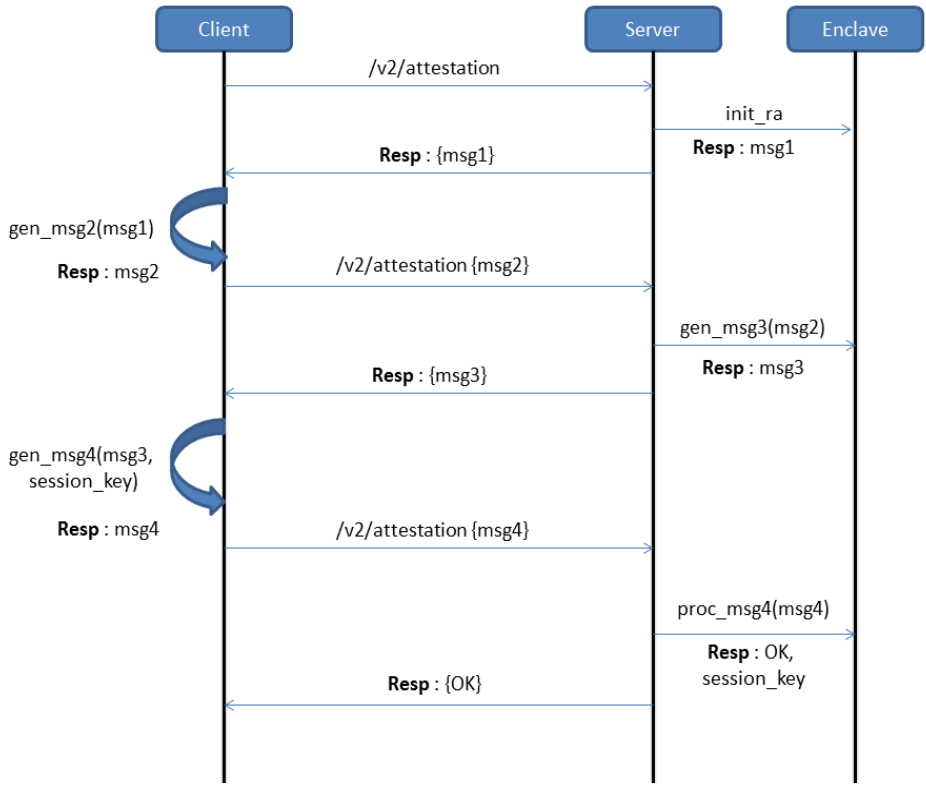

**Figure 4. Remote Attestation REST APIs**

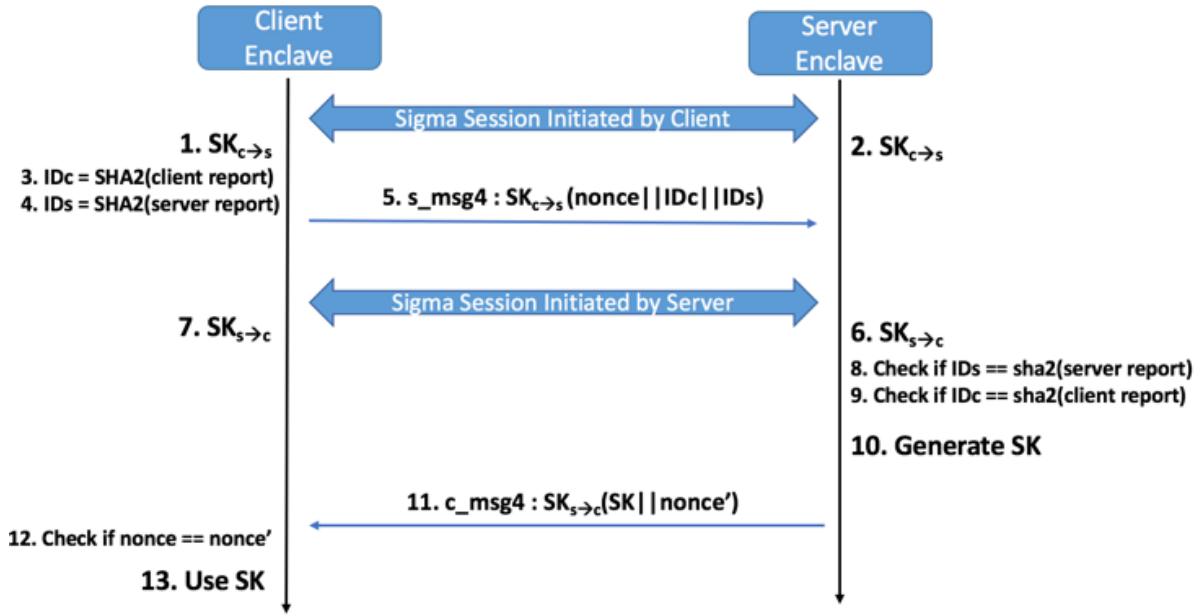

**Figure 5. SK negotiation during Mutual attestation**

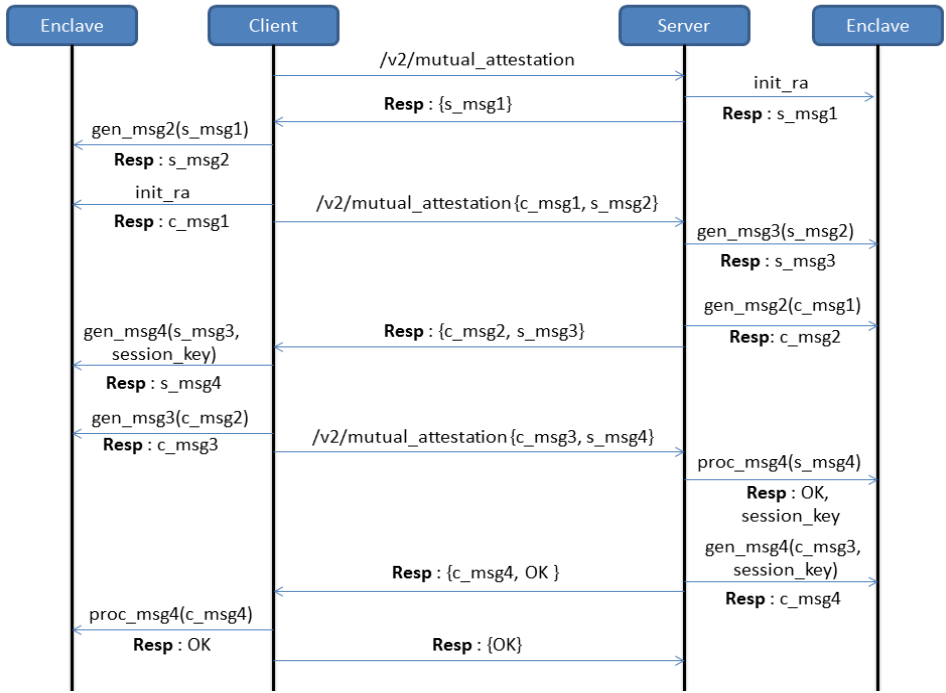

**Figure 6. Mutual Attestation REST APIs**

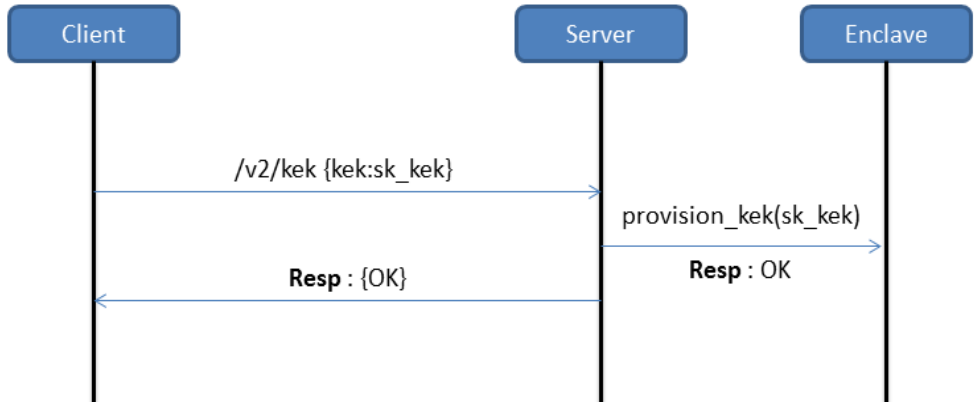

**Figure 7. Master Key (KEK) Provisioning REST APIs**

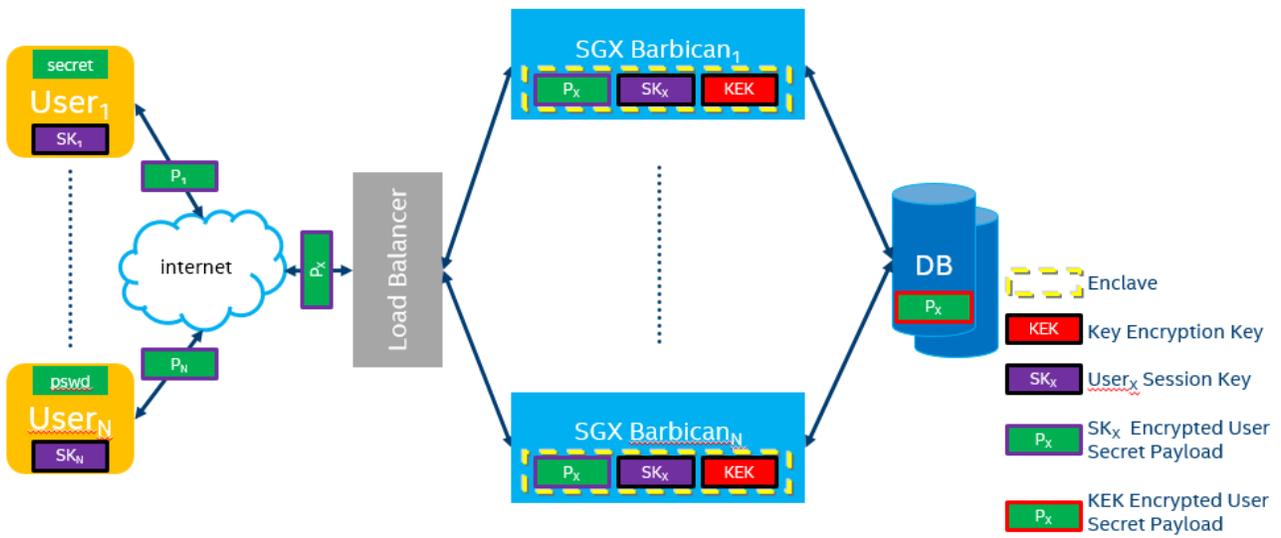

**Figure 8. High Level Architecture**

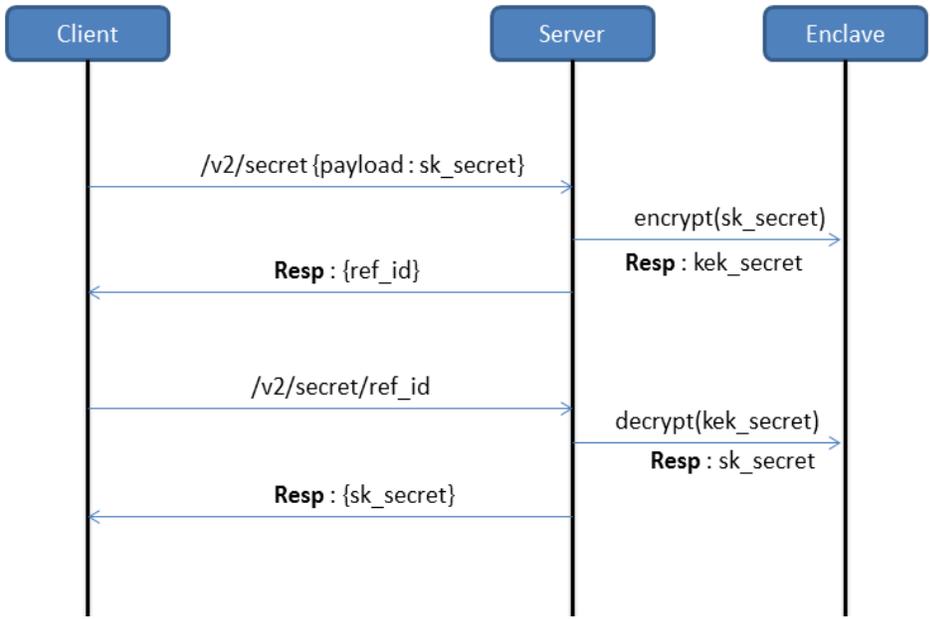

**Figure 9. Secret Management REST APIs**

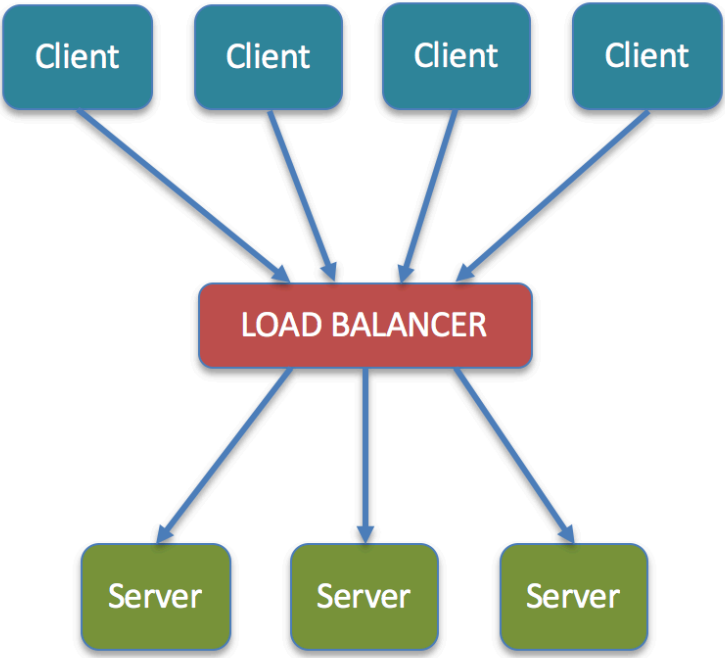

**Figure 10. Cloud Deployment**

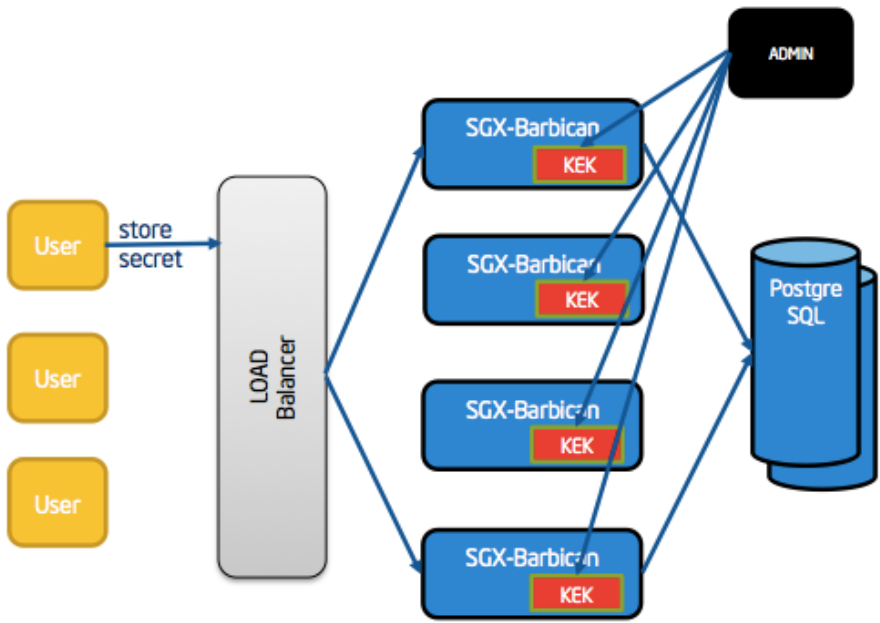

**Figure 11. SGX Barbican Scaled Deployment**

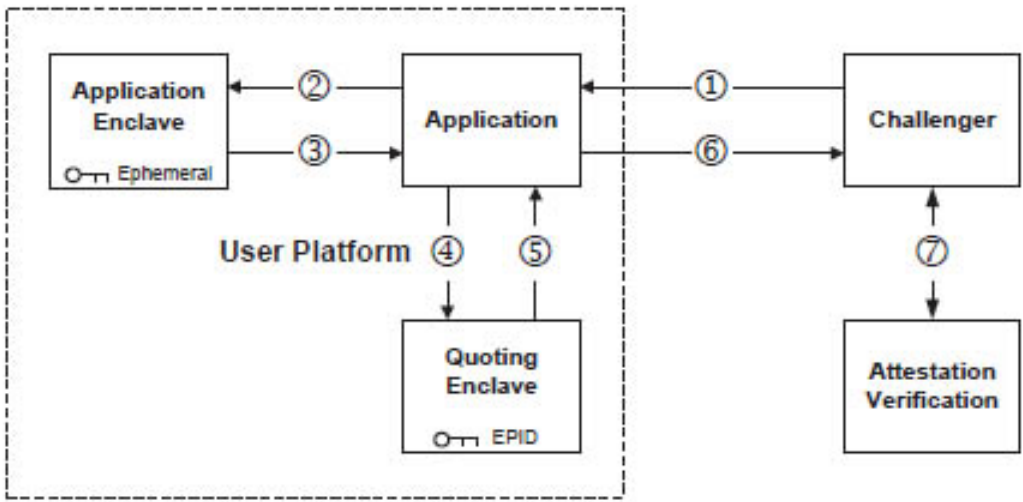

**Figure 12. SGX Remote Attestation**

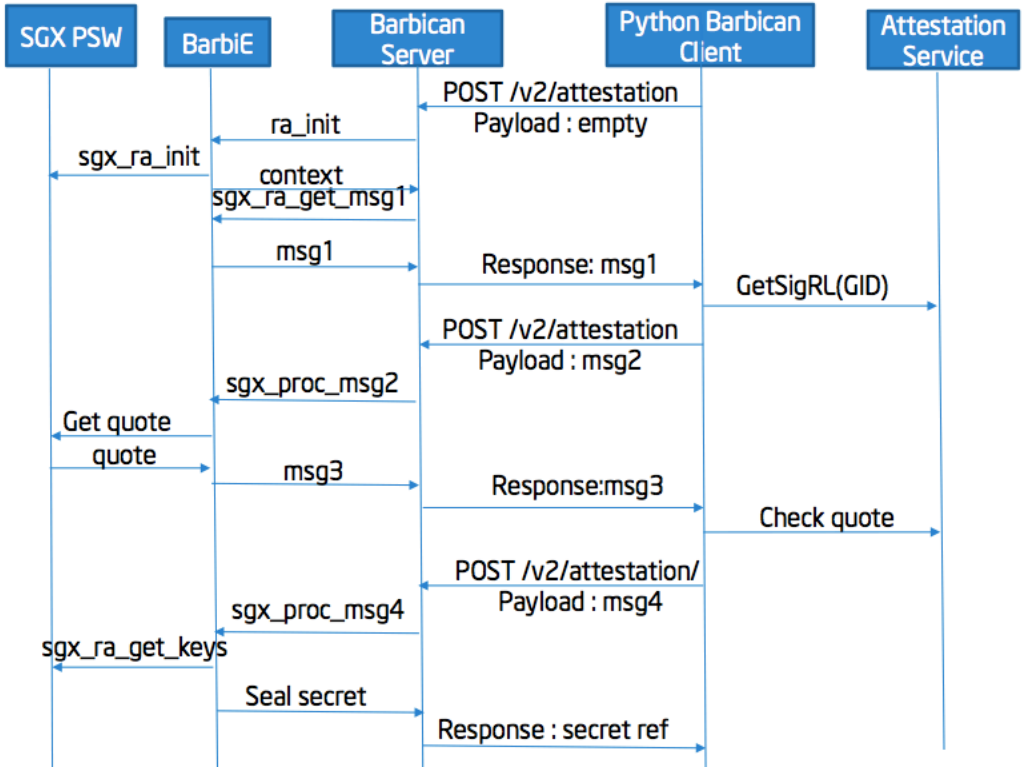

**Figure 13. SGX Remote attestation on scaled deployments**